\documentclass{article}
\usepackage{spconf}
\usepackage{amsmath,graphicx}
\usepackage{color,soul}
\usepackage{mathrsfs}
\usepackage{amssymb}
\usepackage{bbm}
\usepackage{dsfont}
\usepackage{comment}

\newcommand{\mbf}[1]{\mathbf{#1}}
\newcommand{\mbfi}[1]{\mathbf{#1}_i}
\usepackage{arydshln}
\usepackage{algorithm,algcompatible,amsmath}
\usepackage{epstopdf}
\usepackage{graphicx}
\usepackage{multicol}
\usepackage{cite}
\newcommand\norm[1]{\left\lVert#1\right\rVert}
\usepackage{enumitem}

\mathchardef\ordinarycolon\mathcode`\:
\mathcode`\:=\string"8000
\begingroup \catcode`\:=\active
\gdef:{\mathrel{\mathop\ordinarycolon}}
\endgroup

\makeatletter
\renewcommand{\ALG@name}{Table}
\makeatother

\usepackage{tikz}
\usepackage{lipsum}

\newcommand\copyrighttext{%
	\footnotesize 
	Copyright 2018 IEEE. Published in the IEEE 2018 International Conference on Acoustics, Speech, and Signal Processing (ICASSP 2018), scheduled for 15-20 April 2018 in Calgary, Alberta, Canada. Personal use of this material is permitted. However, permission to reprint/republish this material for advertising or promotional purposes or for creating new collective works for resale or redistribution to servers or lists, or to reuse any copyrighted component of this work in other works, must be obtained from the IEEE. Contact: Manager, Copyrights and Permissions / IEEE Service Center / 445 Hoes Lane / P.O. Box 1331 / Piscataway, NJ 08855-1331, USA. Telephone: + Intl. 908-562-3966.}
\def\copyrightnotice{%
	\begin{tikzpicture}[remember picture,overlay]
	\node[anchor=south,yshift=0pt] at (current page.south) {\fbox{\parbox{\dimexpr\textwidth-\fboxsep-\fboxrule\relax}{\copyrighttext}}};
	\end{tikzpicture}%
}

\title{Optimized Transmission for Consensus in Wireless Sensor Networks}
\name{Shahin Khobahi* and Mojtaba Soltanalian \thanks{* Corresponding author (e-mail: \textit{skhoba2@uic.edu}). This work was supported in part by U.S. National Science Foundation Grant CCF-1704401.}
	\address{Department of Electrical and Computer Engineering\\
		University of Illinois at Chicago\\
		Chicago, USA}
}

\begin{document}
\maketitle
\copyrightnotice
\vspace{-0.75cm}
\begin{abstract}
In this paper, we present a consensus-based framework for decentralized estimation of deterministic parameters in wireless sensor networks (WSNs). In particular, we propose an  optimization algorithm to design (possibly complex) sensor gains in order to achieve an estimate of the parameter of interest that is as accurate as possible. The proposed design algorithm employs a cyclic approach capable of handling various sensor gain constraints. In addition, each iteration of the proposed design framework is comprised of the Gram-Schmidt process and power-method like iterations, and as a result, enjoys a low-computational cost.
\end{abstract}
\section{Introduction}
\label{sec:intro}
Wireless sensor networks (WSNs) present significant potential for usage in decentralized detection and estimation due to their many advantageous characteristics such as an inherent distributed structure. While the benefits of digital transmission are well-known, recent research efforts have revealed the superiority of \emph{analog} WSNs in reducing the level of distortion in distributed parameter estimation compared to their \emph{digital} counterparts \cite{gastpar08,gomadam07,marano2007likelihood,marano2013nearest,guerriero2010optimal,khajehnouri2007distributed}. Hence, it is no surprise that analog WSNs have already attracted a considerable attention from researchers---see e.g. \cite{cui07,smith09,banavar12,banavar10,jiang13,jiang14,jiang15}, and the references therein. 

Early works in the context of analog estimation include the study of algorithms for data fusion in both centralized or decentralized scenarios. For instance, the authors in \cite{xiao05} have proposed an average consensus-based decentralized estimation scheme for a network with both fixed and time-varying network topology. In some recent efforts to achieve minimum estimation error, analog amplify-and-forward and phase-shift-and-forward transmission schemes for signal transmission from sensor to fusion center (FC) have been proposed in \cite{cui07},\cite{smith09}, \cite{banavar12}, and \cite{jiang13}, where the sensor gain optimization is usually subject to a total power constraint. Moreover, a distributed parameter estimation algorithm  based on alternating direction method of multipliers (ADMM \cite{boyd11}) has been proposed in \cite{schizas08} and \cite{schizas08-2}. In this paper, we first present an ADMM-based algorithm for estimating the parameter of interest in a decentralized manner. We further formulate the asymptotic variance of the estimation at each node and propose an \textit{efficient} optimization framework that can deal with complex gains of the sensors for an optimized transmission between the nodes to effectively minimize the consensus error variance through the network.

\emph{Graph Notation:} We represent the topology of the WSN by an undirected and connected graph $\mathcal{G}=(\mathcal{E},\mathcal{V})$, consisting of a finite set of vertices $\mathcal{V}=\{1,\dots,n\}$  (also called \textit{nodes}), and a set of edges $\mathcal{E}\subseteq\{\{i,j\}:~i,j\in\mathcal{V}\}$. We denote the edge between node $i$ and $j$ as $\{i,j\}$, which indicates a bidirectional communication between the nodes $i$ and~$j$. We further assume that the sensor connections in $\mathcal{G}$ are time-invariant and the transmissions are always successful.
We define the set of neighbors of node $i$ including itself as $\mathcal{N}_i\triangleq \{j\in \mathcal{V} : ~ \{i,j\} \in \mathcal{E}\}$. The \textit{degree} of the $i$th node is given by $d_{i} = |\mathcal{N}_{i}|$.

\section{System and Fusion Model}
\par
We consider a network with $N$ single-antenna nodes each of which observing an unknown (but deterministic) parameter $\theta \in \mathds{C}$ according to the linear model $z_i = \theta + v_i$ for node~$i$,
where $v_i$ is the observation noise and has the distribution $\mathcal{CN}(0,\sigma_{v,i}^2)$. We further assume that the observation noise is independent from one node to another. Moreover, we assume that the channel state information (CSI) of the network is available at the nodes (at least for the neighbors).
\par
The decentralized estimation scheme operates as follows. The $i$th node amplifies its observation with an adjustable complex gain $a_{i} \in \mathds{C}$ and transmits this amplified observation to its immediate neighbors (i.e., $k \in \mathcal{N}_i$). The received signal at a generic node $k$ from its neighbor node $i$ can be written as $y_{k,i} = h_{k,i}a_{i}z_i + n_{k,i},~ \text{for}~ k\in\mathcal{N}_i$,
where $h_{k,i} \in \mathds{C}$ is the channel coefficient between node $k$ and $i$, and $n_{k,i}$ denotes the transmission noise. Moreover, we assume that the transmission noise is zero-mean  Gaussian noise with variance $\sigma^2_{n}$ and is uncorrelated from one transmission to another. Let $S_i=\{s_1^i,\dots,s_{|\mathcal{N}_i|}^i\}$ denote the ordered sequence of all nodes neighboring the $i$th node. The collection of all observations received at node $i$ can be expressed as $\mbf{z}_i=\mathds{1}\theta+\mbf{v}_i$,
where $\mbf{z}_i = [z_{s^i_1},\dots,z_{s_{|\mathcal{N}_i|}^i}]^T$, and $\mbf{v}_i = [v_{s^i_1},\dots,v_{s_{|\mathcal{N}_i|}^i}]^T$ is the noise vector with covariance $\mbf{V}_i=\mathds{E}\{\mbf{v}_i\mbf{v}_i^H\}=\text{Diag}\{\sigma^2_{v,s^i_1},\dots,\sigma^2_{v,s^i_{|\mathcal{N}_i|}}\}$. Consequently, the received signal vector, this time at the $i$th node, from its neighboring nodes can be expressed as
\begin{equation}
\mbf{y}_{i} = \mbf{H}_i\mbf{D}_i\mbf{z}_i + \mbf{n}_i=
\mbf{H}_i\mbf{a}_i\theta+\underbrace{\mbf{H}_i\mbf{D}_i\mbf{v}_{i}+\mbf{n}_i}_{\triangleq\mbf{w}_i} \label{eq:14},
\end{equation}
where $\mbf{a}_i =[a_{s_1^i},\dots,a_{s^i_{|\mathcal{N}_i|}}]^T$ contains the \emph{sensor gains} to be optimized, $\mbf{D}_i = \text{\text{Diag}}(\mbf{a}_i)$, $\mbf{y_i}=[y_{i,s^i_{1}},\dots,y_{i,s^i_{|\mathcal{N}_i|}}]^T$,  $\mbf{H}_i = \text{\text{Diag}}([h_{i,s_1^i},\dots,h_{i,s^i_{|\mathcal{N}_i|}}]^T)$, and $\mbf{n_i} = [n_{i,s_1^i},\dots,n_{i,s^i_{|\mathcal{N}_i|}}]^T$ is the transmission Gaussian noise vector with covariance $\mbf{R}_{\mbf{n}_{i}}=\mathds{E}\{\mbf{n}_i\mbf{n}_i^H\}=\sigma^2_n\textbf{I}_{|\mathcal{N}_i|}$. Moreover, the covariance of the \emph{combined} Gaussian noise term $\mbf{w}_i$, in \eqref{eq:14} is given by $\mbf{C}_i = \mathds{E}\{\mbf{w}_i\mbf{w}_i^H\}=\mbf{H}_i\mbf{D}_i\mbf{V}_i\mbf{D}_i^H\mbf{H}_i^H + \mbf{R}_{\mbf{n}_i}$. 

A drawback of such an \emph{amplify-and-forward} scheme (governed by variable sensor gains) is that all nodes neighboring the $i$th node will receive the amplified noisy observation $y_{i,k}$,  $k\in\mathcal{N}_i$, and incorporate that single observation into their estimation. Hence, the aggregate global data streams are no longer uncorrelated. In order to further reduce the redundant information in the network, we use the following data compression strategy: Each node starts with initializing a local information value based on \eqref{eq:14}. Namely, the $i$th node calculates $I_i=\mbf{a}_i^H\mbf{H}_i^H\mbf{C}_i^{-1}\mbf{H}_i\mbf{a}_i$ and transmits $I_i$ to its neighboring nodes. Also, note that $I_i$ is an information measure due to the fact that the inverse of $I_i$ provides the variance of the maximum likelihood estimation (MLE) of the parameter.  Next, each node will select one node in its neighborhood with the highest information value and only the selected node will retain the received data from that node, and all other nodes will discard the associated received signal to that node. For instance, consider that the $j$th node has the highest information value among the $i$th node's neighborhood. Then, all nodes $k \in \mathcal{N}_i\backslash\{j\}$ will discard $y_{k,i}$ but the $j$th one. Let $\{\mbf{T}_i\}_{i=1}^{N}$ denote the \emph{row selection matrix} associated with the $i$th node, which points to rows of $\mbf{y}_i$ that are to be discarded. Then, the stacked received data after compression at each node can be described as $\mbf{y}^{\prime}_{i} = \mbf{T}_i\mbf{y}_i$. And, the compressed global observation vector can be written as $\mbf{y}=\mbf{H}\mbf{a}\theta+\mbf{HDv}+\mbf{G}\mbf{n}$, where $\mbf{a}=[a_1,\dots,a_N]$, $\mbf{D}=\text{Diag}\{\mbf{a}\}$, $\mbf{v}=[v_1,\dots,v_N]^T$ whose covariance matrix is $\mbf{V}$, $\mbf{G}=\text{blkdiag}(\{\mbf{T}_i\}_{i=1}^{N})$, $\mbf{H}=[\mbf{T}_1\Omega_1,\dots,\mbf{T}_n\Omega_N]^T$ where $\Omega_k$ is a $|\mathcal{N}_k|\times N$ matrix whose elements are $[\Omega_k]_{ij}=h_{kj}$, if $j\in S_k$ and $j=s^k_i$; otherwise, $[\Omega_k]_{ij}=0$. Also, let the combined global noise term be $\mbf{w}=\mbf{HDv +Gn}$ whose covariance matrix $\mbf{C}=\mathds{E}\{\mbf{ww}^H\}=\mbf{HDVD}^H\mbf{H}^H + \mbf{\Sigma}$ where $\mbf{\Sigma}=\sigma_n^2\mbf{I}_{M}$, and $M=2|\mathcal{E}|-r$ in which $r$ represents the total number of discarded communications. 
 
The ML estimate of $\theta$ given the linear model $\mbf{y}=\mbf{Ha}\theta+\mbf{w}$ can thus be expressed as
\begin{align}
	\hat{\theta}_{ML}&=(\mbf{a}^H\mbf{H}^H\mbf{C}^{-1}\mbf{Ha})^{-1}\mbf{a}^H\mbf{H}^H\mbf{C}^{-1}\mbf{y}\nonumber\\
	&=\left({\sum_{i=1}^{N}\mbfi{a}^H\mbfi{H}^H\mbfi{C}\mbfi{H}\mbfi{a}}\right)^{-1}\sum_{i=1}^{N}\mbfi{a}^H\mbfi{H}^H\mbfi{C}^{-1}\mbfi{y}.\label{eqq:5}
\end{align}
where the ML estimate $\hat{\theta}_{ML}$ is unbiased (\textit{i.e.}, $\mathds{E}\{\hat{\theta}_{ML}\}=\theta$) with variance,
\begin{align}
\label{eqq:6}
	\text{Var}(\hat{\theta}_{ML})&=\left(\sum_{i=1}^{N}\mbfi{a}^H\mbfi{H}^H\mbfi{C}^{-1}\mbfi{H}\mbfi{a}\right)^{-1}\\&=\left(\mbf{a}^H\mbf{H}^H\mbf{C}^{-1}\mbf{Ha}\right)^{-1}.
\end{align}
\subsection{ADMM-Aided Distributed ML Estimation}
The goal now is to facilitate computing \eqref{eqq:5} in a distributed manner. In order to do so, we use an average-consensus scheme based on the alternating direction method of multipliers (ADMM). Particularly, the following ADMM \emph{update equations} were derived in \cite{shi14} to achieve an average consensus in the network:
\begin{align}
y_i^{k+1}&=\frac{1}{1+2\rho|\mathcal{N}_i|}\big(\rho|\mathcal{N}_i|y_i^{k}+\rho\sum_{j\in\mathcal{N}_{i}}y_j^k-\lambda_i^k + x_i\big),\label{eqq:16}\\
\lambda_i^{k+1}& = \lambda_i^{k} + \rho\big(|\mathcal{N}_i|y_i^{k+1}-\sum_{j\in\mathcal{N}_i}y_j^{k+1}\big),\label{eqq:17}
\end{align}
where $y_i^{k+1}$ is the $i$th node's local copy of the global variable (which will eventually converge to the average value of the initial observations, $\bar{x}=(1/n)\sum_{i=1}^{n}x_i$), $x_i$ is the initial observation of node $i$, and $\rho>0$ is an arbitrary constant. As it can be seen from the above update equations, the updates of each node only depend on the \emph{local information}, and the algorithm is hence fully distributed. Next, we use this ADMM-based distributed average consensus scheme to achieve the ML estimate of the parameter. Let $I_i(0) \triangleq \mbfi{a}^H\mbfi{H}^H\mbfi{C}^{-1}\mbfi{H}\mbfi{a}$ be the \textit{information value} at node $i$, and $P_i(0) \triangleq \mbfi{a}^H\mbfi{H}^H\mbfi{C}^{-1}\mbfi{y}$ be the corresponding \textit{state information matrix}. Therefore, each node can (asymptotically) compute the global ML estimate of $\theta$ defined in \eqref{eqq:5} by applying the distributed average consensus steps in \eqref{eqq:16} and \eqref{eqq:17} on the $I_i(0)$ and $P_i(0)$. More precisely, each node updates its information value and the state information matrix according to \eqref{eqq:16}-(\ref{eqq:17}) (by substituting $x_i$ in \eqref{eqq:16} with $I_i(0)$ and $P_i(0)$) and will obtain a local estimate of the parameter of interest at each iteration by computing $\hat{\theta}^i_{ML}(k)=I_i^{-1}(k)P_i(k)$. Due to the fact that, ${I}_c \triangleq \lim_{t\to\infty} I_i(t) = \frac{1}{N}\sum_{i=1}^{N} \mbfi{a}^H\mbfi{H}^H\mbfi{C}^{-1}\mbfi{H}\mbfi{a}$, and ${P}_c  \triangleq \lim_{t\to\infty} P_i(t)  = \frac{1}{N}\sum_{i=1}^{N} \mbfi{a}^H\mbfi{H}^H\mbfi{C}^{-1}\mbfi{y}$, each node will (asymptotically) achieve the ML estimate of the unknown parameter:
\begin{align}
\hat{\theta}_{ML}^i = {I}_c^{-1} {P}_c = \frac{\sum_{i=1}^{N}\mbfi{a}^H\mbfi{H}^H\mbfi{C}^{-1}\mbfi{y}}{\sum_{i=1}^{N}\mbfi{a}^H\mbfi{H}^H\mbfi{C}^{-1}\mbfi{H}\mbfi{a}}.\label{eqq:18}
\end{align}
 In addition, it can be easily shown that the variance of the estimation at each node converges to that of the global ML estimation variance in \eqref{eqq:6}; namely that $\lim_{t\to\infty}\text{Var}(\hat{\theta}^i_{ML}(t))=(\sum_{i=1}^{N}\mbfi{a}^H\mbfi{H}^H\mbfi{C}^{-1}\mbfi{H}\mbfi{a})^{-1}$. In the next section, we devise a low-cost cyclic optimization approach to design the complex gains at each node.
 \vspace{.1cm}
\section{Sensor Gain Optimization}

\label{sec:pagestyle}

Hereafter, we address the problem of designing the (possibly complex) sensor gains $\mbf{a}\in \mathds{C}^{N}$ in order to minimize the variance of the consensus-based estimation given in \eqref{eqq:6}. As it was shown in the previous section, the variance of the estimation at each node asymptotically converges to that of the global ML estimate of the unkown parameter. Our goal here is to minimize $\text{Var}(\hat{\theta}_{ML})$ by considering the  the sensor gain vector $\mbf{a}$ as the optimization variable. In particular, the sensor gain optimization can be formulated as
\begin{align}
\max_{\mbf{a}} &\quad\mbf{a}^H\mbf{H}^H\big(\mbf{HDVD^HH^H + \Sigma}\big)^{-1}\mbf{H}\mbf{a}\label{eq:28}\\
\text{s. t.}&\quad\mbf{a}\in\Omega,\label{eq:29}
\end{align}
where $\Omega$ denotes the search space of the sensor vector $\mbf{a}$. Note that as $\mbf{D}=\text{Diag}\{\mbf{a}\}$, the core matrix of the seemingly quadratic objective in \eqref{eq:28} is a function of sensor gains~$\mbf{a}$. We will show that, by utilizing an over-parametrization approach,  the above optimization problem can be approached via a sequence of quadratic optimization problems.

 Let $\eta = \eta_0 - \mbf{a}^H\mbf{H}^H\big(\mbf{HDVD^HH^H + \Sigma}\big)^{-1}\mbf{H}\mbf{a}$, where $\eta_0$ sufficiently large to keep $\eta$ positive for all $\mbf{a}$ (\textit{e.g.}, $\eta_0 > N||\mbf{H}||^2_F/{\lambda_{min}\{\boldsymbol{\Sigma}\}}$). We will consider the following equivalent optimization problem in lieu of \eqref{eq:28}:
\begin{align}
\min_{\mbf{a}} &\quad\eta\label{eq:30}\\
\text{s. t.}&\quad\mbf{a}\in\Omega\label{eq:31}.
\end{align}
In order to tackle \eqref{eq:30}, let $g(\mbf{{y},{a}})\triangleq\mbf{y}^H\mbf{R}\mbf{y}$, where $\mbf{y}$ is an auxiliary vector variable, and
\begin{equation}
	 \mbf{R}\triangleq\left( \begin{array}{c:c} \rule[-1.5ex]{0pt}{0pt}\quad\eta_0\quad &  \mbf{a}^H\mbf{H}^H \\ \hdashline
	\mbf{Ha} & \rule{0pt}{2ex}    
	\quad\mbf{HDV}\mbf{D}^H\mbf{H}^H+\mbf{\Sigma}\quad \end{array} \right).\label{eq:32}
\end{equation}
Note that $\mbf{e}_1^H\mbf{R}^{-1}\mbf{e}_1=\eta^{-1}$, where $\mbf{e}_1 = (1\, 0 \, \dots \, 0)^T$ is the first standard basis of $ \mathds{R}^{M+1}$. Now consider the optimization problem:
\begin{align}
\min_{\mbf{a,\,y}} &\quad g(\mbf{y},\mbf{a})\label{eq:33}\\
\text{s. t.}&\quad\mbf{y}^H\mbf{e}_1=1,\,\,\mbf{a}\in\Omega.\label{eq:34}
\end{align}
For fixed $\mbf{a}$, the minimizer $\mbf{y}$ of \eqref{eq:33} is given by $\mbf{y}=\left({\mbf{e}_1^H\mbf{R}^{-1}\mbf{e}_1}\right)^{-1}\mbf{R}^{-1}\mbf{e}_1$ (see Result 35 in \cite[p.~354]{stoica05}),
which is in fact a scaled version of the first column of $\mbf{R}^{-1}$ (to satisfy \eqref{eq:34}).  Observe that $\mbf{y}$ is a scaled version of the solution~to~the

\noindent
\begin{minipage}{\linewidth}
	\vspace{-.2cm}
	\begin{algorithm}[H]
		\small
		\caption{The Proposed Sensor Gain Optimization Approach}\label{t:1}
		\textbf{Step 0:} Initialize the auxiliary vector $\mbf{y}$ with a random vector in $\mathds{C}^{M+1}$ such that $y_1=1$. Initialize $\mbf{a}\in\Omega$.\\
		\textbf{Step 1:} Employ the quadratic formulation in \eqref{eq:39}, and particularly the power method-like iterations in \eqref{eq:40} to update the sensor gain vector $\mbf{a}$ (until convergence).\\
		\textbf{Step 2:} Update $\mbf{y}$ using $\mbf{y}=\left({\mbf{e}_1^H\mbf{R}^{-1}\mbf{e}_1}\right)^{-1}\mbf{R}^{-1}\mbf{e}_1$, or by employing the fast approach discussed below \eqref{eq:34}.\\
		\textbf{Step 3:} Repeat steps 1 and 2 until a pre-defined stop criterion is satisfied, \textit{e.g.} $|f(\mbf{a}^{(k)})-f(\mbf{a}^{(k+1)})|\leq\xi$ for some $\xi>0$, where $k$ denotes the outer-loop iteration number, and $f(.)$ is defined as $f(\mbf{a})=\eta$.
	\end{algorithm}
\end{minipage}
\vspace{.2cm}\\
 linear system $\mbf{Ry}=\mbf{e}_1$, and in particular that $\mbf{y}$ is a scaled version of the vector orthogonal to all rows but the first row of $\mbf{R}$. A fast approach to calculate $\mbf{y}$ is therefore to use the Gram-Schmidt process (applied to the rows except the first row of $\mbf{R}$) followed by a scaling.
 \par A more important observation, establishing the \emph{equivalence} of \eqref{eq:30} and \eqref{eq:33}, is that for the minimizer $\mbf{y}$ of \eqref{eq:33} one can easily verify that $g(\mbf{y},\mbf{a})=\eta \nonumber$.
As a result, each step of the cyclic optimization of \eqref{eq:33} with respect to $\mbf{y}$ and $\mbf{a}$ leads to a decrease of $\eta$ (and ultimately convergence, as $\eta$ is lower bounded). In addition, the minimization of \eqref{eq:33} with respect to $\mbf{a}$, and for fixed $\mbf{y}$,  boils down to a quadratic optimization problem. Note that for a feasible $\mbf{y}$ of \eqref{eq:33}, we can partition $\mbf{y}$ as $\mbf{y}^T\triangleq\left(1\,\,\tilde{\mbf{y}}^T\right)$. Therefore, 
\begin{align}
&\mbf{y}^H\mbf{R}\mbf{y}
=\, C_1 +\label{eq:37}\\
&\left( \begin{array}{c} \mbf{a}\nonumber\\
1\rule{0pt}{2ex}    
\end{array} \right)^H
\underbrace{\left( \begin{array}{c:c} \rule[-1.5ex]{0pt}{0pt}\left(\mbf{H}^H\tilde{\mbf{y}}\tilde{\mbf{y}}^H\mbf{H}\right)\odot\mbf{V}\quad &\quad\mbf{H}^H\tilde{\mbf{y}} \\ \hdashline
	\tilde{\mbf{y}}^H\mbf{H} & \rule{0pt}{2ex}    
	0 \end{array} \right)}_{\triangleq\mbf{Q}}\left( \begin{array}{c} \mbf{a}\nonumber\\
1\rule{0pt}{2ex}    
\end{array} \right)\nonumber
\end{align}
where we have used the identity  $\tilde{\mbf{y}}^H\mbf{HDV}\mbf{D}^H\mbf{H}^H\tilde{\mbf{y}}=\mbf{a}^H\left(\left(\mbf{H}^H\tilde{\mbf{y}}\tilde{\mbf{y}}^H\mbf{H}\right)\odot\mbf{V}\right)\mbf{a}$.
\begin{figure*}[t]
	\begin{minipage}[b]{0.48\linewidth}
		\centering
		\centerline{\includegraphics[width=7.1cm]{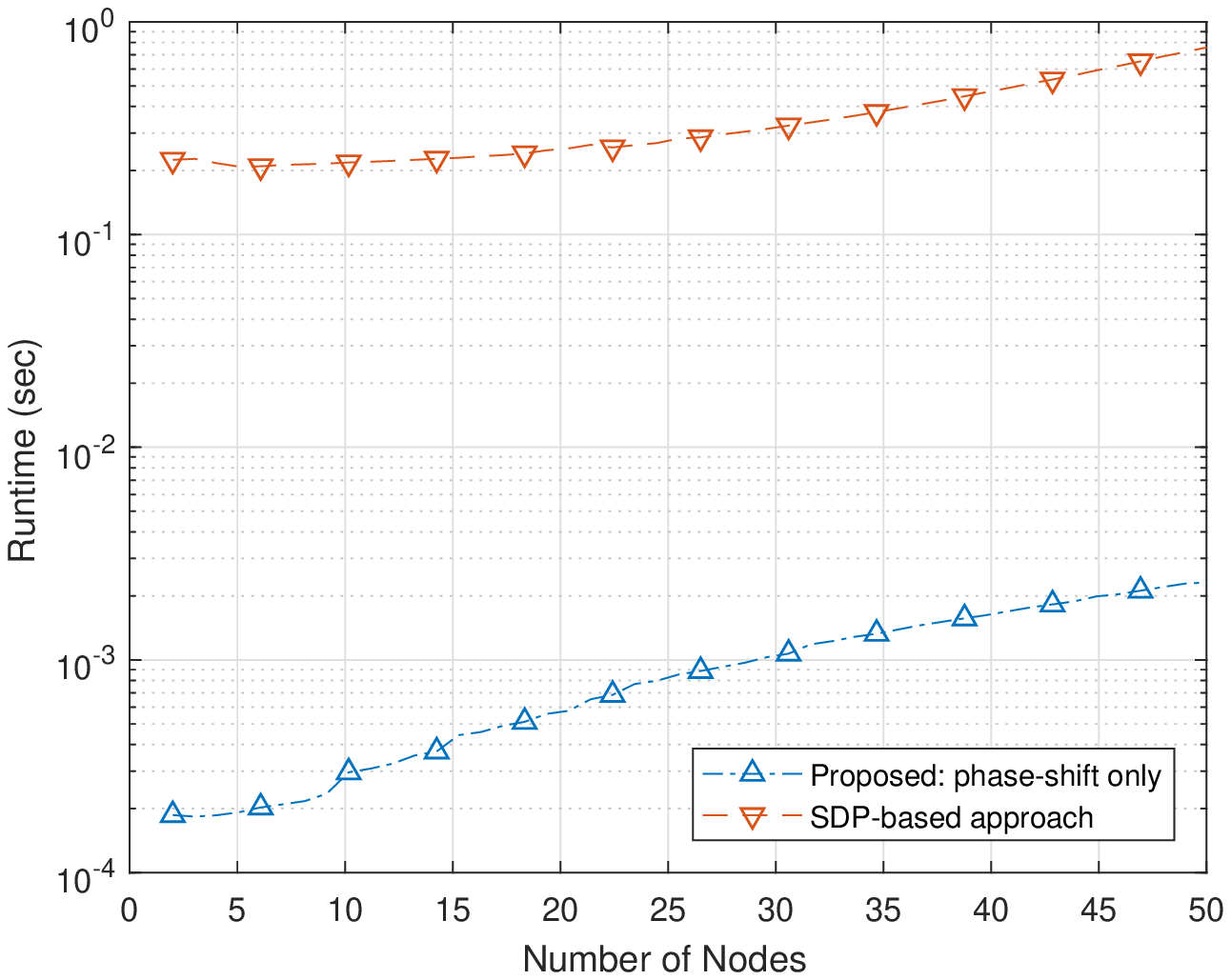}}
		\centerline{(a)}\medskip
	\end{minipage}
	\hfill
	\begin{minipage}[b]{0.48\linewidth}
		\centering
		\centerline{\includegraphics[width=7.1cm]{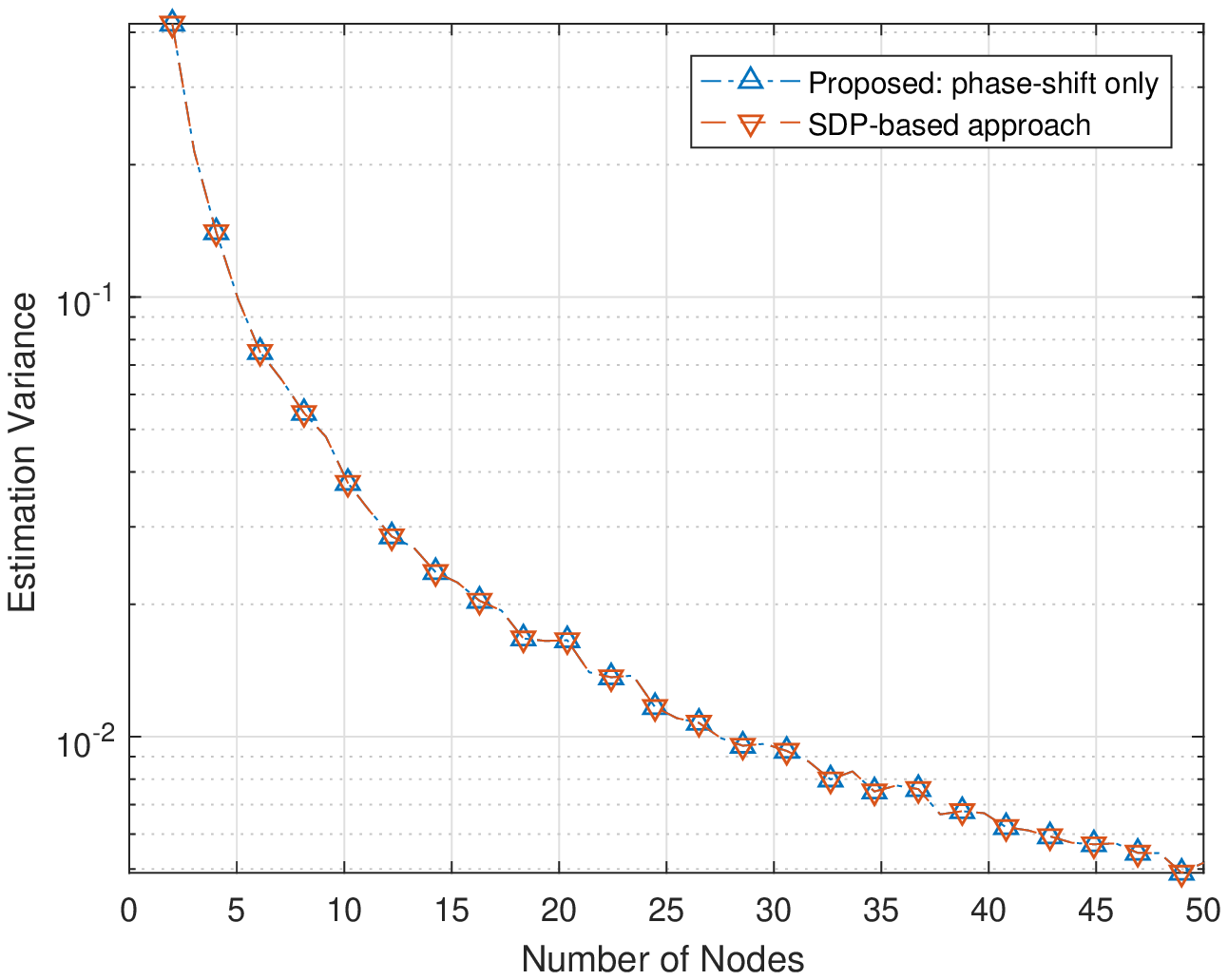}}
		\centerline{(b)}\medskip
	\end{minipage}
	\vspace{-.1cm}
	\caption{Comparison of (a) the runtime and (b) the estimation variance of the proposed method and the SDP-based approach of \cite{jiang13}. The proposed algorithm exhibits significantly lower computational cost, while achieving a similar estimation variance.}
	\label{fig:res}
\end{figure*}
Also note that $C_1 = \eta_0 + \tilde{\mbf{y}}^H\mbf{\Sigma}\tilde{\mbf{y}}$ is invariant with respect to the sensor gain vector $\mbf{a}$. Minimizing \eqref{eq:33} with respect to $\mbf{a}$ can thus be done by considering:
\begin{align}
\max_{\mbf{a}} &\quad \left( \begin{array}{c} \mbf{a}\\
1    
\end{array} \right)^H
\mbf{\tilde{Q}}\,\,\,\,
\left( \begin{array}{c} \mbf{a}\\
1    
\end{array} \right)
\label{eq:39}\\
\text{s. t.}
&\quad\mbf{a}\in\Omega.\nonumber
\end{align}
where $\tilde{\mbf{Q}}\triangleq\lambda\mbf{I}_{M}-\mbf{Q}$ with $\lambda>\lambda_{\text{max}}(\mbf{Q})$, and $\Omega$ is assumed to impose a finite/fixed energy constraint on $\mbf{a}$ (e.g., $||\mbf{a}||_2^2=N$).
Interestingly, a monotonically increasing objective of \eqref{eq:39}, and equivalently a monotonically decreasing objective of~\eqref{eq:33}, can be obtained using the following \textit{power method-like} iterations (see \cite{soltanalian14,naghsh14,soltanalian13,soltanalian142} for details):
\begin{align}
\label{eq:40}
\min_{\mbf{a}^{(t+1)}} &\quad 
\norm{\left( \begin{array}{c} \mbf{a}^{(t+1)}\\
1    
\end{array} \right)^H
-\,\,\,
\tilde{\mbf{Q}}
\left( \begin{array}{c} \mbf{a}^{(t)}\\
1    
\end{array} \right)
}_2\\
\text{s. t.}
&\quad\mbf{a}^{(t+1)}\in\Omega,\nonumber
\end{align}
where $t$ is the iteration number, and $\mbf{a}^{(0)}$ is the current value of $\mbf{a}$. Two useful constrained scenarios for the sensor gain optimization in \eqref{eq:40} are as follows. Let,  $\hat{\mbf{a}}^{(t)}=(\mbf{I}_{M}\,\,\,\mbf{0}_{M\times 1})\,\tilde{\mbf{Q}}
\left( \begin{array}{c} \mbf{a}^{(t)}\\
1    
\end{array} \right)$. The recursions of \eqref{eq:40} for a \textit{finite or fixed energy} scenario can be expressed as $\mbf{a}^{(t+1)}=\left(\sqrt{N}||\hat{\mbf{a}}^{(t)}||_2\right)\hat{\mbf{a}}^{(t)}.$ In addition, for the \textit{phase-shift only} case (i.e. with $|a_i|=1$ for $i=1,\dots,N$), the recursion takes the form $\mbf{a}^{(t+1)}= \text{exp}\left(j\text{arg}\left(\hat{\mbf{a}}^{(t)}\right)\right)$. Note that the latter scenario can be further studied as to a Unimodular Quadratic Program (UQP); see \cite{soltanalian14,naghsh14,soltanalian13,soltanalian142}. Finally, the proposed method is summarized in Table I.
\vspace{-.1cm}
\section{Numerical Results}

\begin{figure}[t]
	\begin{minipage}[b]{\linewidth}
		\centering
		\centerline{\includegraphics[width=7.1cm]{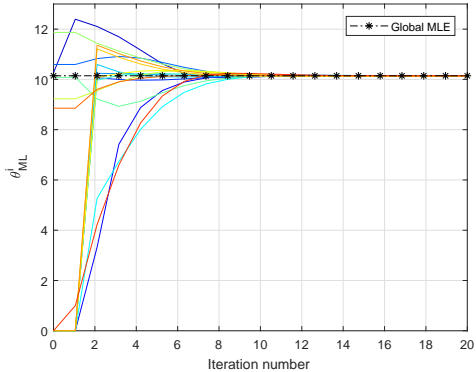}}
		\vspace{-.1cm}
		\caption{Convergence of the ADMM-based ML estimation. }
	\end{minipage}
\end{figure}

\label{sec:typestyle}
In this section, we investigate the performance of our proposed sensor gain optimization algorithm. 
We compare our sensor gain optimization algorithm (Table I) with the state-of-the-art semidefinite programming (SDP) based approach of \cite{jiang13}.  Each measurement is averaged over 300 random channel realizations.  Fig. 1(a) shows a comparison of the computational cost (machine runtime)  between our algorithm and the SDP-based approach in \cite{jiang13}. It is observed from Fig. 1(a) and Fig. 1(b) that although the two algorithms yield similar estimation variance, our proposed optimization algorithm has a significantly lower computational burden. For example, with $N=50$ nodes, one can observe that the runtime of our algorithm is less than $1\%$ of the runtime associated with the SDP-based approach. This is particularly of importance in WSNs since not only the processing resources of the nodes are limited but also that the environment parameters (e.g., the channels) might change and need re-assessments frequently. Hence, it is important for the network to be able to adapt to the new environment \textit{as quickly as possible} with \textit{minimal cost}.  \vspace{.1cm}

Our proposed two-stage algorithm also enables the nodes to obtain the global ML estimation of the parameter based on their \textit{local} information by applying the distributed fusion scheme algorithm described in subsection~2.1. Fig. 2 illustrates the simulation results for this ADMM-based decentralized estimation and the convergence of the proposed decentralized MLE algorithm to that of the global MLE for a network with $N=16$, and $\theta=10$. It can be observed that the local estimate of each node $\hat{\theta}^i_{ML}(k)$ converges to the global MLE of the parameter computed in \eqref{eqq:5}, and a consensus is achieved very quickly.
\vspace{.1cm}
\section{Conclusion}
\label{sec:majhead}
A sensor gain optimization for consensus-based decentralized ML estimation in WSNs was proposed. The presented framework enable the network to quickly converge to the global MLE. Moreover, the proposed sensor gain optimization technique can handle various sensor gain constraints very efficiently---an important feature for large-scale WSNs. 
\vfill\pagebreak
\label{sec:refs}
\bibliographystyle{IEEEbib}
\bibliography{refs,strings}

\end{document}